\documentclass[%
reprint,
amsmath,amssymb,
aps,
]{revtex4-2}

\usepackage{graphicx}
\usepackage{dcolumn}
\usepackage{bm}

\usepackage{tikz}
\usetikzlibrary{arrows.meta}

\renewcommand{\vec}[1]{{\bf #1}}
\newcommand{\xhat}{\hat{x}}
\newcommand{\yhat}{\hat{y}}
\newcommand{\zhat}{\hat{z}}

\begin{document}
	
\preprint{APS/123-QED}

\title{A time-orbiting potential chip trap for cold atoms}

\author{C. A. Sackett}
\email{sackett@virginia.edu}
\affiliation{%
	Department of Physics, University of Virginia,
	Charlottesville, Virginia 22904, USA
}%

\author{J. A. Stickney}
\affiliation{
	Space Dynamics Laboratory, North Logan, UT 84341, USA
}%

\date{\today}

\begin{abstract}
	We present a design for an atom chip trap that uses the time-orbiting potential technique. The design offers several advantages compared to other chip-trap methods. It uses a simple crossed-wire pattern on the chip, along with a rotating bias field. 
	The trap is naturally close to spherically symmetric, and it can be modified to be exactly symmetric in quadratic order of the coordinates. Loading from a magneto-optical trap is facilitated because the trap can be positioned an arbitrary distance from the chip. 
	The fields can be modified to provide a gradient for support against gravity, and the three-dimensional trap can be adiabatically transformed into a two-dimensional guide.
\end{abstract}

\maketitle

Over the past two decades, atom chips have become a critical technology for ultracold atom science 
\cite{Reichel2002,Reichel2011,Keil2016}. An atom chip consists of small current-carrying wires patterned onto a planar substrate. Atoms near the wires experience very large magnetic field gradients, which enables the production of tightly confining magnetic traps with relatively low electrical power consumption. Atom chips are used in many research laboratories, they are the basis for commercial ultra-cold atom systems \cite{Du2004,Farkas2014}, and they have enabled the production of cold atoms in microgravity \cite{Vogel2006,Aveline2020}. 

\begin{figure}[t]
\includegraphics[width=\columnwidth]{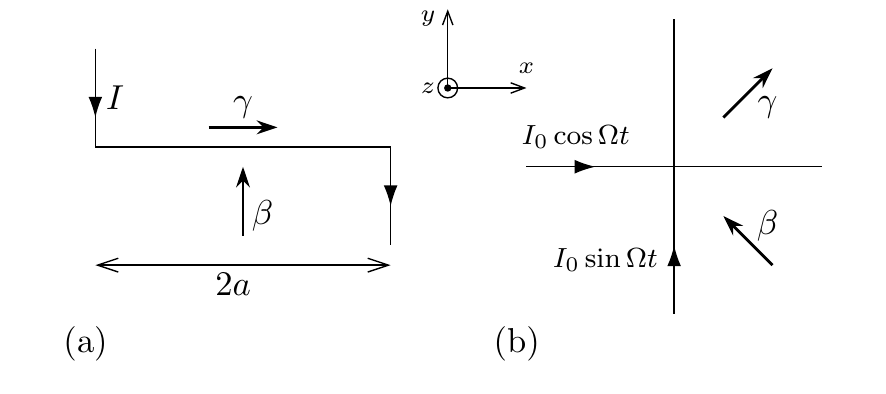}

	\caption{Atom chip configurations. Thin lines represent wires on the chip carrying current $I$, and thicker arrows represent
	uniform field components. (a) Ioffe-Pritchard ``Z trap'' configuration. The $\beta$ field sets the distance of the trap from the chip, while the $\gamma$ field provides a non-zero bias at the trap center. (b) Cross trap configuration. The $\beta$ and $\gamma$ fields play similar roles as in the Z trap, but here the bias fields rotate in sync with the oscillating wire currents. The diagram shows the field and current directions when $\Omega t = \pi/4$.}
	\label{fig1}
\end{figure}

Most implementations of an atom chip use the Ioffe-Pritchard trap configuration \cite{Ketterle1992}. This can be produced, for instance, by a Z-shaped wire as in Fig.~\ref{fig1}(a) \cite{Reichel2002}. Such a ``Z trap'' is suitable for evaporative cooling and has been used to produce quantum degenerate gases in many experiments. It does, however, have some drawbacks that our design aims to redress. First, the distance
of the atoms to the chip is constrained by the Z geometry: 
if the center segment of the Z has length $2a$, then the potential minimum cannot be located further than $1.2a$ from the chip surface due to an inflection point in the field curvature \cite{Sackett2017}. In contrast, for chip distances
much smaller than $a$ the trap confinement is weak along the wire direction, leading to a highly asymmetric trap. 
This problem can be addressed by adding more wires to the chip \cite{Reichel2002}, but in general it is challenging
to implement an approximately spherically symmetric trap over a wide range of chip distances.  

A second drawback of Z traps is that the atoms are necessarily in a state with a non-zero magnetic moment, 
making them
sensitive to background field fluctuations. This can be a limitation for experiments such as atom interferometry \cite{Wang2005} or
entanglement \cite{Riedel2010} 
where the phase evolution of the atoms must be carefully controlled. One way to avoid this problem is
with a Time-Orbiting Potential (TOP) trap \cite{Petrich1995}. Here a uniformly rotating bias field is combined with a static or oscillating gradient field to produce a time-averaged potential that is approximately harmonic. TOP traps are generally insensitive to static or low-frequency field noise since the time average of the atomic moments are zero. TOP traps also permit the use of ac electronic techniques like transformers and resonant circuits, which can simplify the current driver implementation.

TOP traps are typically produced using macroscopic coils and are less confining than chip traps. 
For instance, the atomic Sagnac
interferometer demonstrated in \cite{Moan2020} used a TOP trap produced by cm-scale coils with maximum confinement frequencies of
about 200 Hz \cite{Horne2017} for $^{87}$Rb atoms trapped in the $F=2, m=2$ state. In comparison, chip traps can achieve confinement
frequencies of 1 kHz or more \cite{Reichel2002}. A tighter trap would be useful for applications like the Sagnac interferometer since it would increase the speed of evaporative cooling and thus allow faster operation rates.
The TOP technique has been previously applied with atom chips for a few special uses, 
either to make a torroidal ring trap \cite{Gupta2005} or to reduce roughness in the potential produced by nearby chip wires
\cite{Trebbia2007}, but not to our knowledge to implement a tightly confining trap.

We describe here an atom-chip TOP trap that provides tight confinement with 
no intrinsic geometry scale. The chip wire configuration is shown in Fig.~1(b). 
The concept of this trap is slightly different from that of a conventional TOP trap: Consider first a dc current passing through the $x$ wire of the cross in Fig.~1(b). Adding a bias field $\beta$ in the $+y$ direction produces a line of field zeros running above the $x$ axis. An additional bias field component 
$\gamma$ along $x$ converts this line into a harmonic minimum, which provides confinement along the $y$ and $z$ directions but a uniform potential along $x$. To generate three-dimensional confinement, the 
cross wires are instead driven with
oscillating currents $\cos\Omega t$ and $\sin\Omega t$ while the bias fields rotate in sync.
The shape of the net field is not constant in time, but it approximates a rotating two-dimensional trap.
As long as $\Omega$ is sufficiently large, the
atoms experience the time-averaged field, which results in a three-dimensional trap.

To analyze the system, we set the coordinate origin at the center of the cross. The
fields involved can be expressed as
\begin{align} \label{fields}
	\vec{B}(t) = & \frac{\mu_0 I_0}{2\pi}
	\left[ \frac{y\zhat - z\yhat}{y^2+z^2} \cos\Omega t
	+ \frac{z\xhat - x\zhat}{x^2+z^2} \sin \Omega t \right] \nonumber \\
	& + \beta \left( \yhat \cos \Omega t - \xhat \sin\Omega t \right)
	\nonumber \\
	& + \gamma \left( \xhat \cos \Omega t + \yhat \sin\Omega t \right).
\end{align}
The first line gives the field from the chip wires, which are assumed to be long and thin. Here $I_0$ is the current amplitude, $\mu_0$ is the magnetic constant, and $\Omega$ is the TOP frequency. The second line gives the bias component perpendicular to the wires, with amplitude $\beta$.
The trap center will occur where the chip field and the $\beta$ field cancel, at distance
\begin{equation}
	z_0 \equiv \frac{\mu_0 I_0}{2\pi \beta}.
\end{equation}
We use $z_0$ and $\beta$ as independent variables in the following since $z_0$ is experimentally
significant and the combination leads to relatively simple expressions. We then take implicitly
$I_0 = 2\pi\beta z_0/\mu_0$.
The third line in Eq.~\eqref{fields} is the bias component $\gamma$ that provides a non-zero trap minimum. 
Although the  decomposition shown here is convenient for analysis, the total bias can be implemented as a single rotating field
\begin{equation} \label{bias1}
	\vec{B}_{\text{bias}}(t) = \sqrt{\beta^2+\gamma^2}
	\left[\xhat \cos(\Omega t + \theta) + \yhat \sin(\Omega t + \theta)\right]
\end{equation}
with phase $\theta = \tan^{-1}(\beta/\gamma)$ relative to the chip currents.

To characterize the trap, we Taylor expand the field components around the trap center, with $\zeta \equiv z-z_0$.
The field magnitude is
\begin{align}
	B(t) & = \sqrt{B_x^2+B_y^2+B_z^2} \nonumber \\
	& \approx \gamma + 
 \frac{\beta^2}{2\gamma z_0^2} \left(x^2 \cos^2\Omega t
	+ y^2 \sin^2 \Omega t + \zeta^2\right) \nonumber \\
	& \qquad -\frac{2\beta}{z_0^2} (x^2+xy-y^2)\sin\Omega t \cos\Omega t 
\end{align}
to second order in the coordinates. Time averaging yields the effective potential
\begin{equation} \label{quad}
	V(\vec{r}) = \mu \langle B \rangle = \mu\gamma + \frac{\mu\beta^2}{4\gamma z_0^2}\left(\rho^2+2\zeta^2\right)
\end{equation}
where $\mu$ is the magnetic moment of the atomic state and $\rho^2= x^2+y^2$. The potential is confining and cylindrically symmetric, with harmonic oscillation frequencies
\begin{equation} \label{freqs}
	\omega_\rho = \sqrt{\frac{\mu \beta^2}{2 m \gamma z_0^2}}, \quad
	\omega_z = \sqrt{2}\omega_\rho
\end{equation}
for atomic mass $m$. 

In comparison, a Z trap with chip distance $z_0 \ll a$ has oscillation frequencies
\begin{equation}
	\omega_y^{(Z)} = \omega_z^{(Z)} =  \sqrt{\frac{\mu \beta^2}{m \gamma z_0^2}},
	\quad
	\omega_x^{(Z)} = \frac{2z_0^2}{a^2} \omega_z^{(Z)} 
\end{equation}
where $\beta$ and $\gamma$ are again the transverse and longitudinal bias fields. Here we see that the net curvature
$\omega_x^2 + \omega_y^2 + \omega_z^2$ is the same for both traps (neglecting $z_0^4/a^4$), while 
$(\omega_x \omega_y \omega_z)^{1/3}$ is larger in the cross trap by a factor of $(a/z_0)^{2/3}$. The
density of the trapped atom cloud is set by the geometric mean, making it most relevant for efficient evaporative cooling and many other applications.

\begin{figure}
	\includegraphics[width=0.9\columnwidth]{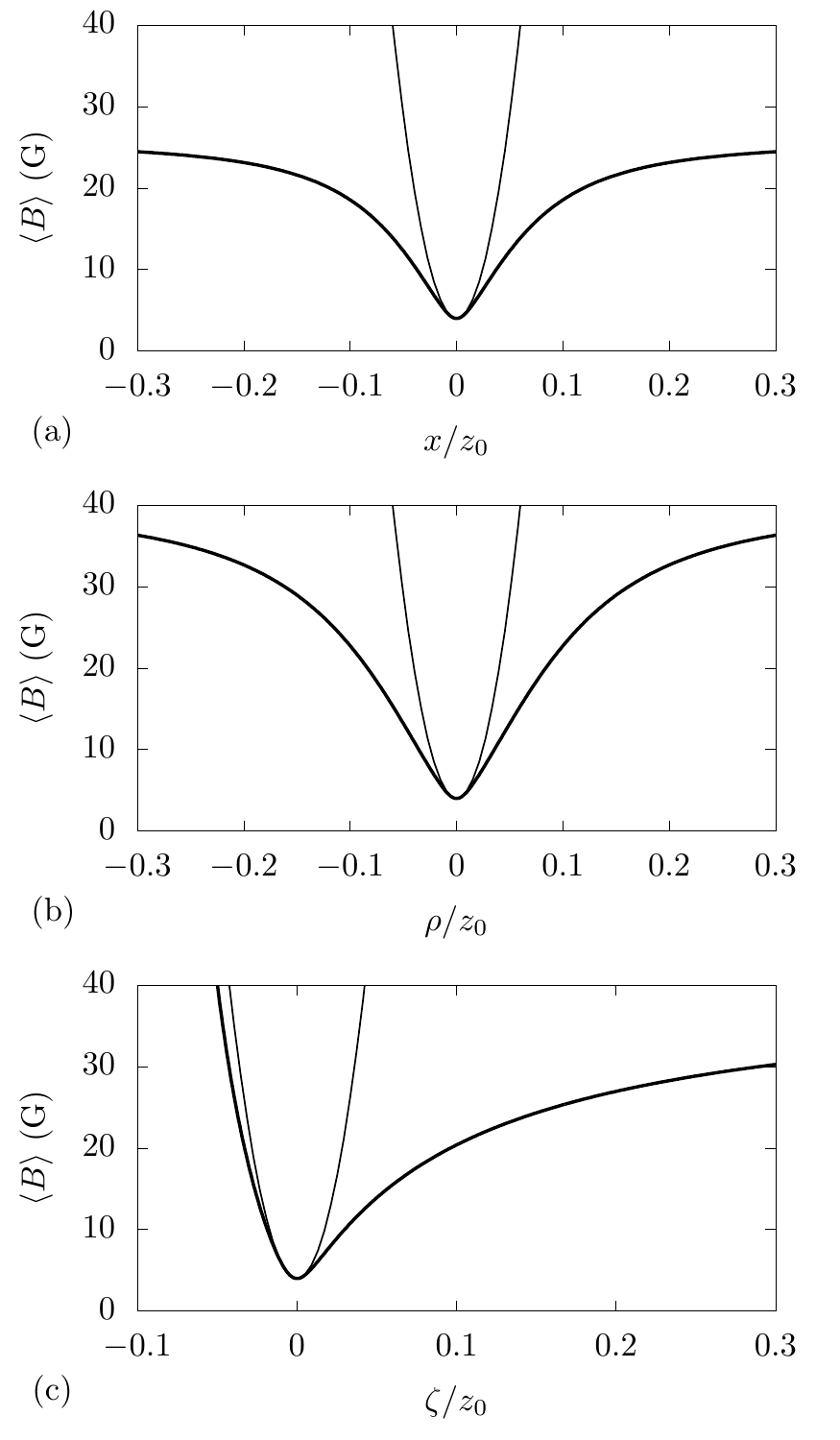}
	\caption{ \label{fig2} Trapping potential for cross trap, for parameters $I_0 = 20$~A, $\beta = 40$~G, and $\gamma = 4$~G. These provide a potential minimum at $z_0 = 1$~mm. The heavier curves show the time-averaged magnetic field magnitude and the lighter curves are the quadratic approximation of Eq.~\protect\eqref{quad}.  
	(a) Plot of the average field and quadratic approximation along the $x$ axis at $z = z_0$. (b) Plot of the average field and quadratic approximation along the line $x=y$,
	for $\rho = \sqrt{x^2+y^2}$ and with $z = z_0$. (c) Plot of the average field and quadratic approximation
	along the $z$ axis, with $\zeta = z-z_0$.}
\end{figure}

Figure \ref{fig2} compares the numerically calculated trap potential to the harmonic
approximation derived above. As to be expected, the confining potential is harmonic only very near
the trap center. 
It is possible to extend the analytical calculation
to higher orders and extract the leading anharmonic terms. With the aid of symbolic math software, we find the fourth-order expansion
\begin{align}
	\langle B \rangle &\approx \gamma + \frac{\beta^2}{4\gamma z_0^2}\Bigg\{\rho^2 + 2z^2 -\frac{2}{z_0}\big(\rho^2 z + z^3\big) \nonumber \\
	& -\frac{1}{16 z_0^2}\Bigg[\left(20+\frac{3\beta^2}{\gamma^2}\right) \rho^4 
	+ \left(28+\frac{3\beta^2}{\gamma^2}\right)x^2y^2\nonumber\\
	& \qquad\qquad +\frac{8\beta^2}{\gamma^2}\big(xy^3 - x^3y\big) + 8\left(\frac{\beta^2}{\gamma^2}-4\right)\rho^2z^2
	\nonumber\\
	& \qquad\qquad+8\left(\frac{\beta^2}{\gamma^2}-12\right)z^4
	\Bigg]\Bigg\},
\end{align}
The anharmonic terms become
important for coordinate excursions on the order of $z_0$ or $\gamma z_0/\beta$, whichever is smaller.

The preceding results confirm that there is no intrinsic geometrical length scale for the
cross chip TOP trap, since $z_0$ can be made as small or large as desired simply by adjusting the field and current amplitudes. In practice, however,
the range of $z_0$ will be constrained on the large side by the length $L$ of the cross wires. 
The impact of finite $L$ will depend on how current is delivered to the chip. If the current enters via long 
lead wires perpendicular to the chip, the dominant effect is that the leads contribute a field parallel to
the $\beta$ field, which moves the trap minimum closer to the chip and makes the trap more confining. 
If $\beta$ is reduced to keep $z_0$ 
constant, there is a modest reduction in the confinement frequencies. For $L/z_0 > 4$, 
the reduction is less than 10\%. 
The range of $z_0$ is limited on the small side by the width $w$ of the chip wires, since the thin-wire
approximation will fail. If the wires are modeled as flat strips, we find that 
as $z_0$ is reduced, the trap minimum moves closer to the chip than $z_0$ and the confinement
becomes weaker. Both $\Delta z/z_0$ and $\Delta \omega/\omega$ remain less than 10\% down to $z_0 = w$.

\begin{figure}
	\includegraphics[width=0.9\columnwidth]{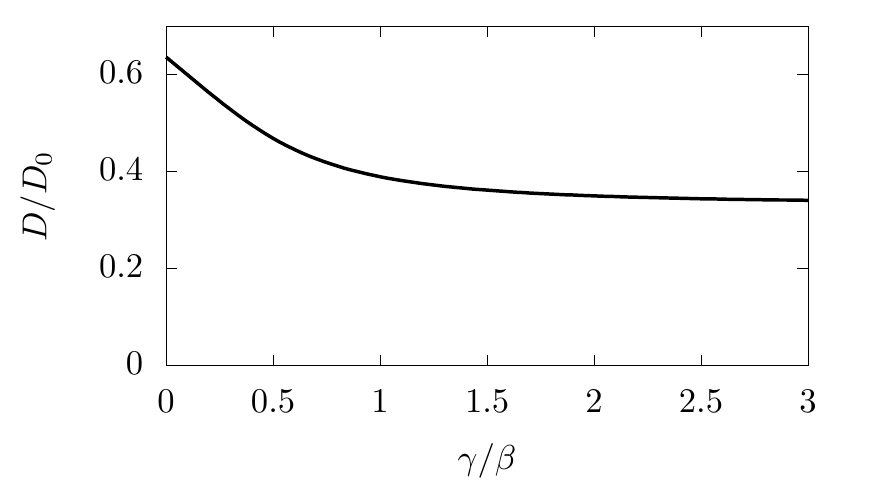}
	\caption{ \label{fig3} Trap depth $D$ for the cross trap, where $D_0 = \sqrt{\beta^2 + \gamma^2}-\gamma$. For large $\gamma/\beta$, the depth approaches $D_0/3 = \beta^2/6\gamma$, 
		and for $\gamma/\beta \rightarrow 0$, the depth approaches $2D_0/\pi = 2\beta/\pi$.}
\end{figure}

Unlike a conventional TOP trap \cite{Petrich1995}, the cross chip trap has no field zero, so there is no ``circle of death'' limiting the trap depth. Instead the depth $D$ is set by the time-averaged field above the wires far from the origin. The depth cannot be expressed as a simple
analytic function, but it is of order $D_0 \equiv \sqrt{\beta^2 + \gamma^2}-\gamma$. A numerical 
calculation of the depth is shown in Fig.~\ref{fig3}.

Applications such as atom interferometry can make use of a weakly confining trap, in which case it is necessary to compensate for gravity.
The cross trap can achieve this by changing the relationship between the
chip fields and the bias fields. A convenient parametrization is via a phase $\phi$ in the 
$\gamma$ field of Eq.~\eqref{fields}, making it 
$\gamma[\xhat \cos(\Omega t+\phi) + \yhat \sin (\Omega t + \phi)]$. In terms of the total bias field of Eq.~\eqref{bias1}, this corresponds to correlated shifts in amplitude and phase
$|\vec{B}_{\text{bias}}| \rightarrow \sqrt{\beta^2+\gamma^2+2\gamma\beta\sin\phi}$ and 
$\theta \rightarrow \tan^{-1}[(\beta+\gamma\sin\phi)/(\gamma\cos\phi)]$. Re-evaluating the time-averaged field
to second order yields
\begin{align} \label{phifield}
	\langle B \rangle = & \,\, \gamma + \beta \sin\phi \frac{\zeta}{z_0} 
	+ \left(\frac{\beta^2}{4\gamma} + \frac{1}{2} \beta \sin\phi\right)\frac{\rho^2}{z_0^2} \nonumber \\
&	+ \left(\frac{\beta^2}{2\gamma}\cos^2\phi - \beta\sin\phi\right) \frac{\zeta^2}{z_0^2}.
\end{align} 
The term linear in $\zeta$ can compensate for gravity in the $z$ direction.

We can also use this approach to model a case where the $\gamma$ field rotation rate is different from $\Omega$,
by setting $\phi = \Delta t$ for constant $\Delta$.
We then have $\langle \sin\phi \rangle \rightarrow 0$ and 
$\langle \cos^2\phi \rangle \rightarrow 1/2$, leading to a spherically symmetric trap with isotropic frequency
$\omega^2 = \mu\beta^2/(2m\gamma z_0^2)$. One way to achieve this is with $\Delta = -\Omega$, corresponding to 
a static field $\gamma$ pointing in any direction parallel to the chip. Use of a static field, however, would re-introduce sensitivity to dc background fields.

As an example of a potential application, we describe an atom chip capable of capturing atoms from a MOT located several mm from the chip, and then compressing the atoms to a trap with confinement frequencies above 1 kHz for evaporative cooling. We consider a chip fabricated from 100-$\mu$m thick direct-bonded copper on an aluminum-nitride substrate \cite{Squires2011}. The side of the chip facing the atoms is patterned to produce cross wires that are 
100~$\mu$m wide.  The opposite side is has a matching cross pattern with wires
3 mm wide. The chip size $L$ is 3~cm and the chip thickness is 1~mm.  
The wider cross is used to produce a distant trap for loading. 
Using a current amplitude $I_0 = 75$~A, bias fields
$\beta = 20$~G, $\gamma = 2$~G, and a phase $\phi = 0.85$~rad, the resulting trap is 7~mm from the chip.
For $^{87}$Rb atoms in the $F=2, m_F=2$, Zeeman state where $\mu$ is equal to the Bohr magneton, this trap provides support against gravity and confinement frequencies
$\omega_\rho \approx 2\pi\times 18$~Hz and $\omega_z \approx 2\pi\times 13$~Hz, with a 
trap depth of 12~G $\approx $ 800~$\mu$K. These are appropriate values for direct loading from a MOT
\cite{Squires2016}.
The total power consumption on the chip is about 10~W, which is well within the capacity of 
this type of substrate \cite{Squires2011}.  

Once the trap is loaded, current through the wide cross can be adiabatically decreased, 
which reduces $z_0$ and compresses the trap. Once the atoms are within a few mm of the chip,
the current is adiabatically shunted to the thin cross, supporting smaller $z_0$. A current of 5 A and
bias fields $\beta = 40$~G, $\gamma = 2$~G would generate a trap 0.25~mm from the 
chip surface with $\omega_\rho \approx 2\pi\times 1$~kHz and $\omega_z \approx 2\pi\times 
1.4$~kHz. This makes a suitable trap for rapid evaporative cooling. Power dissipation on the
chip would be about 1~W. If a two-layer chip as described here is undesirable, another way to support a wide range of $z_0$ values is with tapered wires whose widths decrease as they approach the cross center.
We note that the trap considered here is far enough from the chip that roughness of the
wire and other surfaces effects are unlikely to be significant \cite{Trebbia2007,Harber2003}.

An important question for a TOP chip trap is the value of the oscillation frequency $\Omega$. 
The frequency must be large compared to the highest confinement frequency of the trap, $\omega_m$,
so
that atom motion is negligible during the TOP period $2\pi/\Omega$. The frequency must also 
be small compared to the Larmor frequency $\approx \mu_B \gamma/\hbar$, so that the TOP fields do
not drive spin transitions. Typical TOP frequencies are on the order of 10~kHz, while
typical confinement frequencies are on the order of 100~Hz. Because the atom-chip trap presented
here can achieve confinement frequencies above 1~kHz, it may be necessary to use a
correspondingly greater TOP frequency.

The minimum usable ratio of TOP frequency to confinement frequency has not, to our knowledge, 
been previously explored. The lowest ratio we find in the literature uses
$\Omega/\omega_m \approx 20$ \cite{Hodby2000}. The trap described
by Horne and Sackett \cite{Horne2017} uses a TOP field rotating at 10~kHz and a maximum confinement frequency of 200~Hz,
but the plane of of the TOP field precesses at 1~kHz.  The potential experienced by the atoms is significantly modulated at the 1 kHz frequency without observable effects, suggesting that $\Omega/\omega_m \gtrsim 5$ may be sufficient. 
These results indicate that a 1.4~kHz chip trap as described above could use a TOP frequency
below 30~kHz, and perhaps as low as 7~kHz. TOP traps operating at 20 kHz have been demonstrated \cite{Kozuma1999}, so we expect the drive requirements here to be achievable. At a bias field of 2~G, the Larmor frequency for
$^{87}$Rb is 1.4~MHz, so the parameters proposed here do not approach the high frequency limit.

Another technical concern is how the chip current sources could be implemented. Since the two chip wires
intersect, 
it is necessary either for the two driver circuits to float with respect to ground, or for each driver to be balanced
so that the center of the cross is at a common ground potential. Either of these solutions can be readily achieved using isolation transformers, which are efficient and stable at
frequencies of order 10~kHz \cite{Horowitz1989}.

A final noteworthy feature of the cross TOP configuration is that the three-dimensional trap can 
be adiabatically converted to a two-dimensional guide. 
This can be achieved by reducing
the current through one of the wires to zero along with the corresponding $\beta$ field component. 
For a guide along the $x$ axis, the resulting field is
\begin{align} \label{guide}
	\vec{B}(t) = & \beta \cos\Omega t\left[\yhat +
	\frac{z_0( y\zhat - z\yhat)}{y^2+z^2} 
	\right]	\nonumber \\
	& + \gamma \left( \xhat \cos \Omega t + \yhat \sin\Omega t \right),
\end{align}
with still $z_0 = \mu_0 I_0/2\pi\beta$ for chip current amplitude $I_0$. The time-averaged field has 
the form
\begin{equation}
	\langle B \rangle = \gamma + \frac{\beta^2}{4\gamma z_0^2}\left(y^2 + \frac{3}{4}\zeta^2\right)
\end{equation}
with $\zeta = z-z_0$,
and thus provides harmonic confinement with $\omega_y^2 = \mu\beta^2/(2\gamma z_0^2)$ and
$\omega_z = (\sqrt{3}/2)\omega_y$. 
For example,
if $\beta = 40$~G, $\gamma = 2$~G and $I = 5$~A as in the trap previously considered, the guide distance
remains at 0.25~mm and the confinement frequencies for $^{87}$Rb are about 1~kHz and 800~Hz. Power dissipation
on the chip is reduced by a factor of two compared to the equivalent trap. The guide potential can again be
modified to support gravity 
by introducing a phase $\phi$ to the $\gamma$ field as in Eq.~\eqref{phifield}, resulting
in
\begin{align}
	\langle{B}\rangle = & \ \gamma + \frac{1}{2}\beta  \sin\phi \frac{\zeta}{z_0}
	+ \left(\frac{\beta^2}{4\gamma} + \frac{1}{2}\beta\sin\phi\right)\frac{y^2}{z_0^2} \nonumber \\
&	+ \left[\frac{\beta^2}{16\gamma}\left(1+2\cos^2\phi\right) - \frac{1}{2}\beta\sin\phi\right]\frac{\zeta^2}{z_0^2}.
\end{align}
Linear guides are useful for many applications involving atom
transport \cite{Keil2016}, including atom interferometry \cite{Mueller1999,Wang2005}.

In summary, the cross TOP trap provides a chip-based trap with confinement comparable or better than that of typical Ioffe-Pritchard configurations. The
confinement is naturally cylindrically symmetric and can be readily modified to be spherically symmetric and to provide support against gravity. 
The trap center can be positioned further
from the chip than possible with conventional approaches, and the same chip geometry can provide a two-dimensional atom guide.
 We expect that these features will make the cross TOP useful for a variety of applications. One example is the atomic Sagnac interferometer of \cite{Moan2020}, 
where the cross trap could significantly simplify the apparatus and allow faster production of Bose condensates, thus increasing the sensing 
bandwidth. For this purpose, the cylindrical symmetry of the trap is critical.
We are also exploring how the approach could be extended to produce bias fields with the chip itself, and thereby remove the need for external coils.
By such means, we hope this method will
facilitate the use of ultracold atom techniques in practical applications.

\begin{acknowledgments}
	This work was supported by DARPA (Award No.\ FA9453-19-1-0007). The authors thank  M. Beydler, E. Imhof, B. Kasch, E. Moan, and E. Salim for helpful advice and conversations. 
\end{acknowledgments}

\vfill

\bibliographystyle{apsrev4-2}

\end{document}